\documentclass[letterpaper]{article} 
\usepackage{aaai2026}  
\usepackage{times}  
\usepackage{helvet}  
\usepackage{courier}  
\usepackage[hyphens]{url}  
\usepackage{graphicx} 
\usepackage{natbib}  
\usepackage{caption} 
\usepackage[most]{tcolorbox}
\tcbuselibrary{breakable}
\usepackage{tabularx}
\usepackage{booktabs}
\usepackage{colortbl}
\usepackage{url} 
\usepackage{array}
\usepackage{makecell}
\usepackage{amssymb}
\usepackage{algorithm}
\usepackage{algorithmic}
\usepackage{booktabs}
\usepackage{multirow}

\usepackage{amsmath}
\usepackage{newfloat}
\usepackage{listings}
\DeclareCaptionStyle{ruled}{labelfont=normalfont,labelsep=colon,strut=off} 
\floatstyle{ruled}
\newfloat{listing}{tb}{lst}{}
\floatname{listing}{Listing}
\usepackage[switch]{lineno} 
\title{ComLQ: Benchmarking Complex Logical Queries in Information Retrieval}

\author {
    Ganlin Xu\textsuperscript{\rm 1},
    Zhitao Yin\textsuperscript{\rm 1},
    Linghao Zhang\textsuperscript{\rm 1},
    Jiaqing Liang\textsuperscript{\rm 1},
    Weijia Lu\textsuperscript{\rm 2},
    Xiaodong Zhang\textsuperscript{\rm 2},
    Zhifei Yang\textsuperscript{\rm 2},
    Sihang Jiang\textsuperscript{\rm 3},
    Deqing Yang\textsuperscript{\rm 1}\thanks{Corresponding author},
}
\affiliations {
    \textsuperscript{\rm 1}School of Data Science, Fudan University, Shanghai, China\\
    \textsuperscript{\rm 2}United Automotive Electronic Systems, Shanghai, China\\
    \textsuperscript{\rm 3}College of Computer Science and Artificial Intelligence, Fudan University, Shanghai, China\\
   glxu24@m.fudan.edu.cn, yangdeqing@fudan.edu.cn
}

\usepackage{bibentry}

\begin{document}

\maketitle

\begin{abstract}
Information retrieval (IR) systems play a critical role in navigating information overload across various applications. Existing IR benchmarks primarily focus on simple queries that are semantically analogous to single- and multi-hop relations, overlooking \emph{complex logical queries} involving first-order logic operations such as conjunction ($\land$), disjunction ($\lor$), and negation ($\lnot$). 
Thus, these benchmarks can not be used to sufficiently evaluate the performance of IR models on complex queries in real-world scenarios. To address this problem, we propose a novel method leveraging large language models (LLMs) to construct a new IR dataset \textbf{ComLQ} for \textbf{Com}plex \textbf{L}ogical \textbf{Q}ueries, which comprises 2,909 queries and 11,251 candidate passages. A key challenge in constructing the dataset lies in capturing the underlying logical structures within unstructured text. Therefore, by designing the subgraph-guided prompt with the subgraph indicator, an LLM (such as GPT-4o) is guided to generate queries with specific logical structures based on selected passages. All query-passage pairs in ComLQ are ensured \emph{structure conformity} and \emph{evidence distribution} through expert annotation. To better evaluate whether retrievers can handle queries with negation, we further propose a new evaluation metric, \textbf{Log-Scaled Negation Consistency} (\textbf{LSNC@$K$}). As a supplement to standard relevance-based metrics (such as nDCG and mAP), LSNC@$K$ measures whether top-$K$ retrieved passages violate negation conditions in queries. Our experimental results under zero-shot settings demonstrate existing retrieval models' limited performance on complex logical queries, especially on queries with negation, exposing their inferior capabilities of modeling exclusion. In summary, our ComLQ offers a comprehensive and fine-grained exploration, paving the way for future research on complex logical queries in IR.
\end{abstract}
\begin{links}
\link{Code and Datasets}{https://github.com/xgl-git/ComLQR-main}
\end{links}
\section{Introduction}
\begin{figure}[h]
    \centering
    \includegraphics[width=1\linewidth]{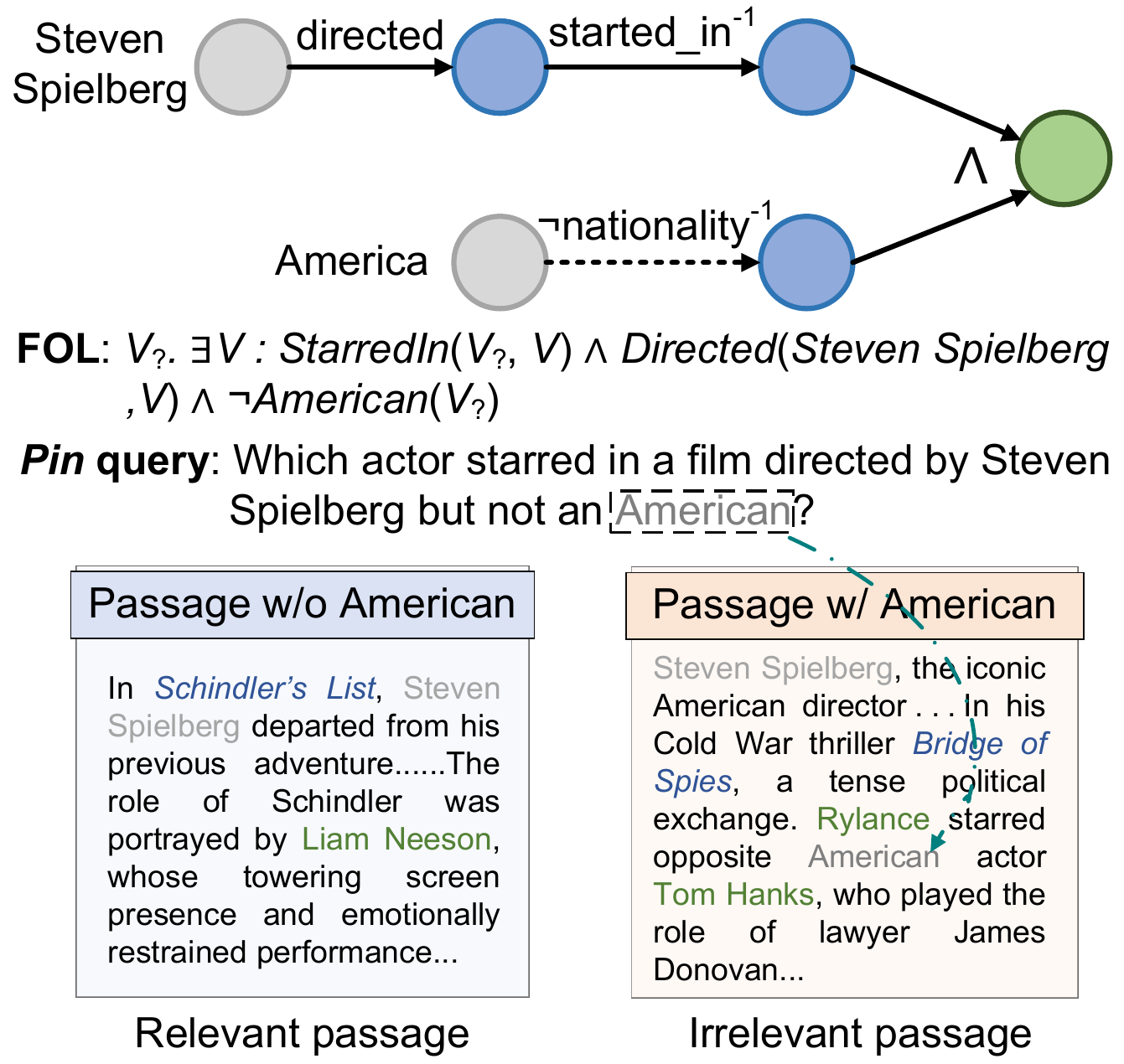}
    \caption{Given a complex logical query of \emph{pin} type, the state-of-the-art retriever InteR tends to retrieve the passages containing the keyword `American', which, however, is not relevant to the query.}\label{figure1}
\end{figure}

Information retrieval (IR) systems, as a cornerstone in addressing information overload, have been widely adopted in various AI applications, including recommendation systems \cite{10.1145/3637528.3671458}, question answering \cite{karpukhin-etal-2020-dense}. A typical IR system retrieves relevant documents or passages from a designated corpus in response to user queries \cite{10.1145/3637870}. Traditional IR benchmarks such as MS-MARCO \cite{DBLP:conf/nips/NguyenRSGTMD16}, TREC \cite{DBLP:conf/trec/CraswellMMYC20}, and BEIR \cite{NEURIPS-DATASETS-AND-BENCHMARKS2021_65b9eea6}, predominantly focused on relatively simple queries that are semantically analogous to single- and multi-hop relations requiring limited compositional inference, which fail to satisfy complex retrieval needs of users in real-world applications. According to our analysis using GPT-4o to classify queries, over 93\% of queries in existing IR benchmarks are such simple queries.
Yet, real user queries are often more complex and involve compositional logical reasoning, highlighting the limitations of these IR benchmarks in capturing the diversity and complexity of user queries. \\
\indent Complex logical queries can be represented with first-order logic (FOL) that involves logical operations such as conjunction ($\land$), disjunction ($\lor$), negation ($\neg$) and existential quantifier ($\exists$) \cite{NEURIPS2020_e43739bb, DBLP:conf/iclr/RenHL20}. For example, given a complex logical query of \emph{pin} type in Figure~\ref{figure1} ``\emph{Which actor starred in a film directed by Steven Spielberg but not an American?}" where \emph{p, i} and \emph{n} represent \emph{projection}, \emph{intersection} and \emph{negation}, respectively \cite{NEURIPS2020_e43739bb, DBLP:conf/iclr/RenHL20}, it can be formalized in FOL as: 
\begin{align*}
V_{?}.~ \exists V:~ & \textit{StarredIn}(V_{?}, V) \land~ \textit{Directed}(\text{Steven Spielberg}, V) \\
                    & \land~ \neg \textit{American}(V_{?}).
\end{align*}
Although complex logical queries received considerable attention in the community of knowledge base question answering (KBQA) \cite{fangetal2024complex, ji-etal-2024-retrieval}, they remain underrepresented in existing IR benchmarks\footnote{Unlike KBQA's operations on structured triples with explicit entities and relations, IR systems execute set operations such as intersection, union, projection, and negation directly over unstructured text. Please refer to Section 2.1 for more details.}. Compared to simple queries, handling complex logical queries requires the precise localization of information resources and thus is conducive to achieving logical reasoning.
These queries involve intricate retrieval intents, which cannot be handled through simple word co-occurrence and thus pose significant challenges for current IR systems. Figure~\ref{figure1} shows that, for the given query, the state-of-the-art retriever InteR \cite{feng-etal-2024-synergistic} relying on word co-occurrence favors the irrelevant passage containing the keyword `American' rather than the relevant passage without the keyword, revealing that they fail to understand query intents correctly \cite{xu-etal-2025-logical}.

To overcome the underrepresentation of \textbf{Com}plex \textbf{L}ogical \textbf{Q}ueries in existing IR benchmarks, we introduce a novel IR dataset, namely \textbf{ComLQ} in this paper, to provide a comprehensive and fine-grained exploration for complex logical queries, which includes 2,909 queries and 11,251 candidate passages. A key challenge in constructing the dataset lies in capturing the underlying logical structures within unstructured text. To address the problem and ensure the quality of the dataset, we design a data synthesis process applicable across diverse domains. Specifically, we first select passages from the existing corpus. To automatically acquire queries with specific logical structures, we then design a subgraph-guided prompt to request an LLM to generate queries based on selected passages, where a key component \emph{subgraph indicator} guides the LLM to learn subgraph patterns associated with different query types. Finally, experienced annotators review each generated query-passage pair concerning the following two criteria. $i$) \textbf{Structure conformity} ensures that queries not strictly conforming to the intended query structure are filtered out. $ii$) \textbf{Evidence distribution} ensures that for queries generated from multiple passages, supporting evidence is indeed distributed across those passages.
Furthermore, to better evaluate whether retrievers can handle queries with negation in ComLQ, we propose a new evaluation metric \textbf{Log-Scaled Negation Consistency} (\textbf{LSNC@$K$}), which measures the extent to which the retrieved top-$K$ passages violate negation conditions in queries.\\
\indent We conduct experiments on a wide range of retrieval models on our ComLQ under zero-shot settings, of which the significant findings include: \textbf{\emph{I}}) None of the experimented retrievers can consistently outperform other methods across all query types, highlighting the need for approaches tailored to different logical structures. \textbf{\emph{II}})
All retrievers exhibit consistent performance degradation as query complexity increases, revealing their limitations in capturing and reasoning over intricate logical structures. \textbf{\emph{III}})  The order of logical operations in query structures notably affects retrievers' performance, since their performance on projection-then-intersection queries (e.g., pi, pin and pni) is worse than that on intersection-then-projection queries (e.g, ip and inp). \textbf{\emph{IV}}) All experimented retrievers exhibit low LSNC@100 scores on queries with negation, revealing a critical gap in IR models' ability to model exclusion.

The main contributions of this paper include:

1. We introduce a new IR dataset \textbf{ComLQ} for benchmarking complex logical queries in IR. To the best of our knowledge, ComLQ is the first IR dataset offering a comprehensive and fine-grained exploration for complex logical queries.

2. We propose an effective method to automatically acquire queries with specific logical structures based on LLMs' generation, where a subgraph-guided prompt involving the \emph{subgraph indicator} is specially designed to guide LLMs to learn subgraph patterns associated with different query types.

3. To evaluate whether retrievers handle queries with negation, we propose a novel metric, LSNC@$K$, to supplement existing relevance-based metrics (nDCG and mAP).

4. We conducted extensive experiments on a wide range of retrieval models, and found that existing retrievers exhibit significant limitations on complex logical queries, with consistently low scores on the proposed metric LSNC@100, revealing their inability to model exclusion. Our findings in this paper pave the way for future research on complex logical queries in IR.
\begin{table*}[h]
	\centering
        \footnotesize
	\begin{tabular}{lccccccccccccccc}
		\toprule
		Model type & 1p & 2p & 3p & 2i & 3i & pi & ip & 2u & up & 2in & 3in & inp & pin & pni &total\\
		\midrule
		Query number& 176 & 227 & 189 & 207 & 267 & 168 & 240 & 258 & 193 & 278 & 271 & 173 & 140 & 122 & 2,909 \\
		\bottomrule
	\end{tabular}
    \caption{The statistics of all query types in ComLQ.}\label{tab1}
\end{table*}

\section{Related Work}
\subsection{Complex Logical Queries}
In recent years, reasoning over single-hop and multi-hop relational data \cite{yang-etal-2018-hotpotqa, LIN2023253} has made remarkable advances. In addition, subsequent research has explored more complex logical structures that involve unobserved edges, multiple entities, and variable interactions \cite{NEURIPS2023_6174c67b}. In this paper, we focus on conjunctive logical queries \cite{NEURIPS2018_ef50c335}, a subclass of first-order logic queries characterized by existential quantifier $\exists$ and conjunction $\land$. Conjunctive logical queries require a set of anchor entities, $\mathcal{V}$, a unique target entity $V_?$ representing the answer to the query, and a set of existential quantified variables $V_1, \cdots, V_m$, and are defined as the conjunction of literals $e_1, \cdots, e_n$:
\begin{equation}
    q = V_?, \exists V_1, \cdots, V_m : e_1 \land e_2 \land \cdots \land e_n, \label{eq1}
\end{equation}
where $e_i$ is an edge involving variable nodes and anchor nodes, satisfying $e_i = r(v_j, V_k)$, $V_k \in \{V_?, V_1, \cdots, V_m\}$, $v_j \in \mathcal{V}$, $r \in \mathcal{R}$, or $e_i = r(V_j, V_k)$, $V_j, V_k \in \{V_?, V_1, \cdots, V_m\}$, $j \neq k$, $r \in \mathcal{R}$. $\mathcal{R}$ is the set of relations defined in the knowledge base (KB).

Although prior work on complex logical queries has predominantly focused on knowledge base question answering (KBQA) \cite{fang-etal-2024-complex, 10.1145/3637528.3671869}, these approaches typically operate over structured triples with a predefined entity-relation schema, where reasoning is performed along explicit relation paths. In contrast, our retrieval setting shifts the focus from path-based reasoning on structured triplets to executing set-theoretic operations (such as projection, intersection, union and negation) directly over unstructured text. This requires retrievers to identify and combine relevant spans from natural language passages that jointly satisfy the query’s logical intent. As a result, our task demands not only semantic understanding but also the recovery of implicit logical structures in open-domain text.

\subsection{Information Retrieval Benchmarks}
Traditionally, the evaluation of information retrieval (IR) systems has relied on standardized benchmarks and well-established evaluation metrics. Numerous datasets evaluated IR systems from Wikipedia \cite{leeetal2019latent}, web queries \cite{bajaj2018msmarcohumangenerated} and biomedical questions \cite{10.1145/345508.345577}. Recently, several benchmarks combine multiple datasets and evaluate retrieval or embedding models across different domains and use cases, such as BEIR \cite{NEURIPSDATASETSANDBENCHMARKS202165b9eea6}, and MTEB \cite{muennighoff-etal-2023-mteb}. Besides, increasing efforts have focused on vertical domains, such as climate science \cite{schimanski-etal-2024-climretrieve}, the legal domain \cite{su-etal-2024-stard} and academic scholarship \cite{ajith-etal-2024-litsearch}. These domains are typically characterized by dense specialized terminology and a strong reliance on domain-specific knowledge, placing greater demands on the professional adaptability of IR systems.

Despite these advancements, current IR benchmarks still fall short in providing a comprehensive and fine-grained evaluation for complex logical queries involving projection, intersection, union and negation, as well as various combinations of these operations. NegConstraint \cite{xu-etal-2025-logical} only focuses on negative-constraint queries, which are similar to the queries with negation in our work. Multi-hop benchmarks, such as HotpotQA \cite{yang-etal-2018-hotpotqa}, neglect other essential logical operations, including intersection, union and negation, which are explicitly represented in our ComLQ. Therefore, they represent only a subpart (or subset) of ComLQ's broader scope. The TREC Complex Answer Retrieval (CAR) track \cite{dietz2017trec} focuses on generating and restructuring long textual answers, whereas our ComLQ emphasizes the structural formulation of queries themselves. \citet{NEURIPS2023_78f9c04b} introduce complex scientific questions which, however, are long-form questions containing multiple simple sentences, and excluding complex logical structures.

\begin{figure}[t]
    \centering
    \includegraphics[width=\linewidth]{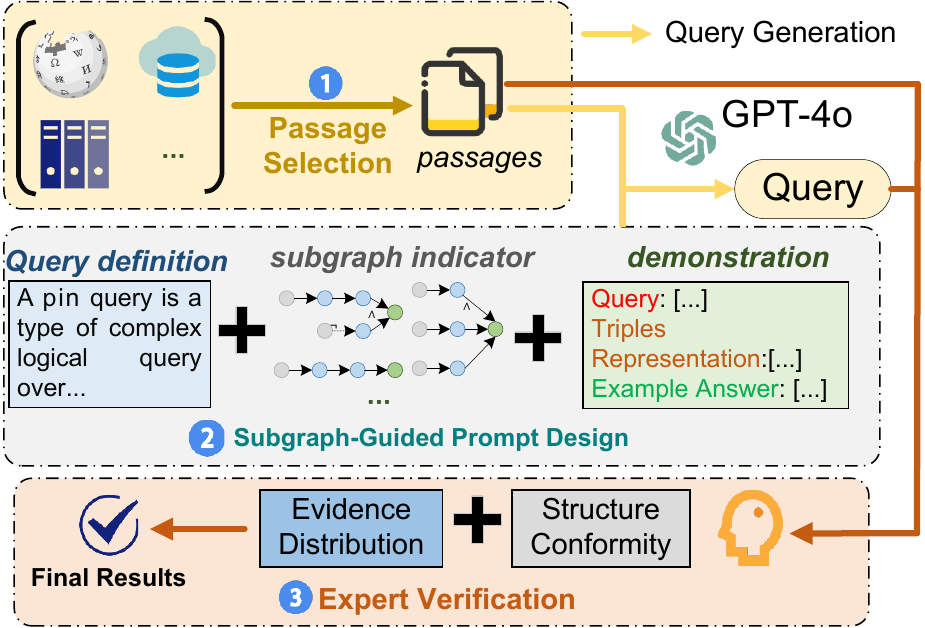}
    \caption{The data synthesis process of ComLQ.}\label{figure2}
\end{figure}
\section{Dataset Construction}
\subsection{Dataset Overview}
In this paper, we adopt the standard definition of complex logical queries from \cite{NEURIPS2020_e43739bb, DBLP:conf/iclr/RenHL20}, consisting of 9 query types without negation (denoted as \emph{1p/2p/3p/2i/3i/pi/ip/2u/up}) and 5 query types with negation (denoted as \emph{2in/3in/inp/pin/pni}), where \emph{p, i, u} and \emph{n} represent \emph{projection}, \emph{intersection}, \emph{union} and \emph{negation} in query structure, respectively. All data in our ComLQ are organized in a standard format (corpus, queries, qrels\footnote{Qrels (query relevance judgments) are ground-truth annotations indicating which passages are relevant to specific queries.}) akin to the BEIR benchmark \cite{NEURIPSDATASETSANDBENCHMARKS202165b9eea6}. The average length of a query in ComLQ is 19.95 words, and the average length of a passage is 112.74 words. Table~\ref{tab1} lists the query number of each type, where queries with negation account for 33.8\%, ensuring balanced distributions to evaluate retrieval performance on modeling exclusion. Query examples and structures are introduced in Appendix A. We request the annotators to score each query-passage pair using a 3-point grading scale (0-2):
\begin{itemize}
\item[$\bullet$]  \textbf{Level-0}. The passage is irrelevant (fully mismatched) to the query. 
\item[$\bullet$]  \textbf{Level-1}. The passage is partially relevant to the query and partly satisfies the query's information needs. That is, if supporting evidence for the query is distributed across two or three passages, each passage is labeled as Level-1.
\item[$\bullet$]  \textbf{Level-2}. The passage content is customized to satisfy the information needs of the query and precisely contains the answer to the query. If supporting evidence for the query is entirely contained within a single passage, that passage is labeled as Level-2.
\end{itemize}

\subsection{Data Synthesis}

Figure~\ref{figure2} shows the data synthesis process of ComLQ, which is applicable across diverse domains. We first select passages from the existing corpus, for which we use the standard Wikipedia dump as the knowledge source in this paper. Then, we design a \emph{subgraph-guided prompt} to request LLMs to generate queries conforming to specific query structures based on the selected passages. Although queries are generated by LLMs, the process is structure-constrained and passage-grounded, ensuring that each query is anchored in real content and follows well-defined logical structures. Finally, in the \emph{expert verification}, three experienced annotators carefully examine each query-passage pair by two criteria: \emph{structure conformity} and \emph{evidence distribution}.

\begin{figure}[t]
    \centering
    \includegraphics[width=0.75\linewidth]{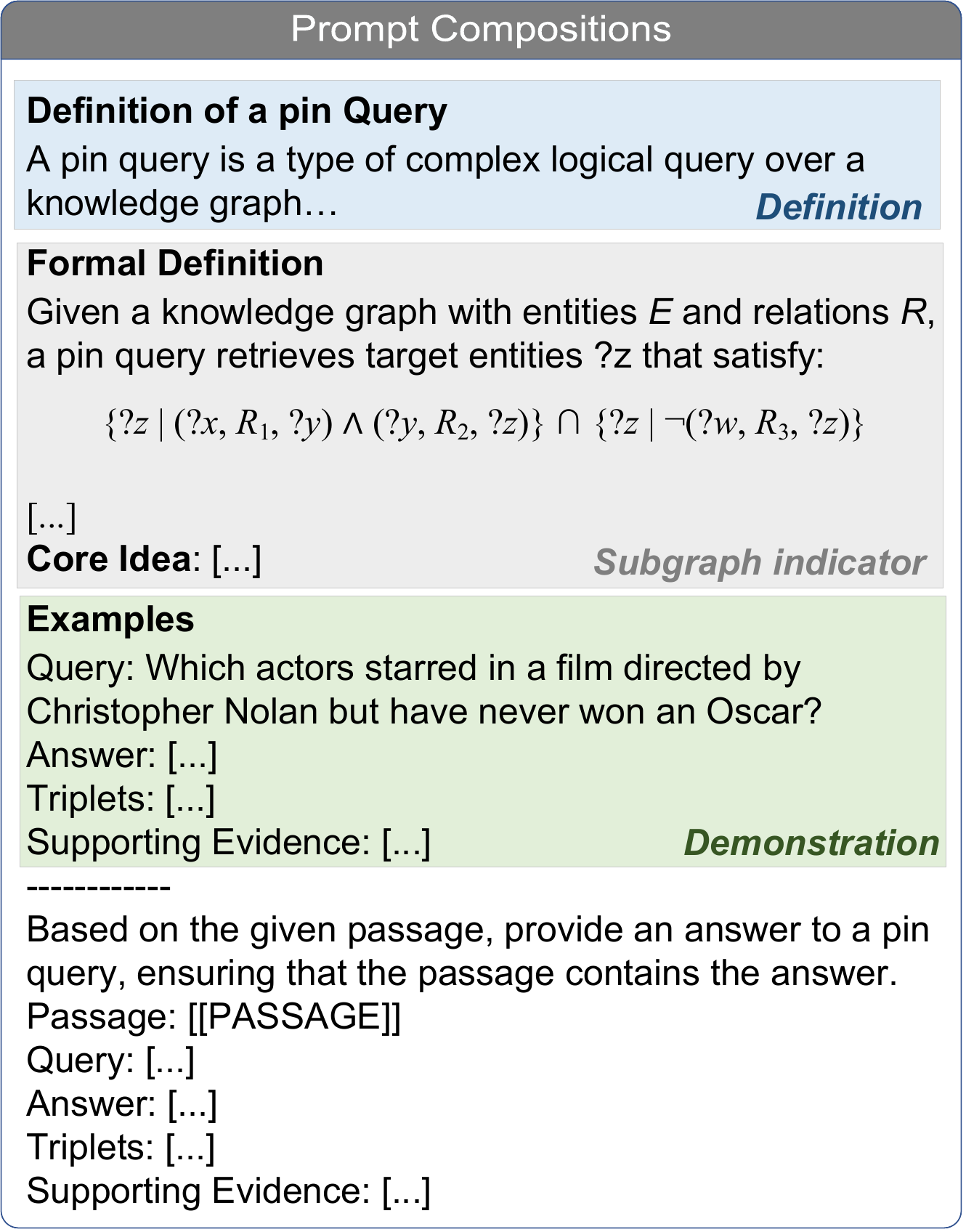}
    \caption{Prompt compositions for generating \emph{pni} query sample.}\label{figure3}
\end{figure}

\subsubsection{Passage Selection}
According to \cite{xu-etal-2025-logical}, we use the Wikipedia dump as the knowledge source in this paper, from which 20 million passages are obtained by segmenting 5 million titled articles. We select one or multiple passages from the same article each time to generate corresponding queries. When selecting multiple passages for query generation, we ensure that all passages are topically relevant to the query. That is, queries require reasoning across passages sharing the same title, which aims at evaluating the retrievers' ability to aggregate relevant information from multiple passages.
\begin{table}[t]
	\centering
        \scriptsize
        
    \centering
	\begin{tabular}{|c|m{5.5cm}|c|}
		\hline
		 \multirow{11}{*}{\textbf{\makecell{Positive \\ Query}}} & \makecell{\textbf{Query}: Which actor starred in a film directed 
         by Steven \\ Spielberg but not an American?} & \multirow{11}{*}{\checkmark} \\
		\cline{2-2}
		&\makecell[l]{\textbf{Triplets}: \\\hspace{1em}\textbf{Projection}: \\\hspace{2em}(StevenSpielberg, directed, ?y)\\ \hspace{2em}(?y, starred by, ?z) \\\hspace{1em}\textbf{Negation}: \\ \hspace{2em}$\neg$(?z, nationality, America)\\ \hspace{1em}\textbf{Intersection}: \\ \hspace{2em}(StevenSpielberg, directed, ?y) $\land$ (?y,  starred by, \\ \hspace{2em}?z) $\land$ $\neg$(?z, nationality,  America)} & \\
		\cline{1-3}
		\multirow{9}{*}{\textbf{\makecell{Negative \\ Query}}} & \makecell{\textbf{Query}: Which microprocessors were
        considered for the \\ original IBM  PC but
        were not used in its final design?} & \multirow{9}{*}{\texttimes} \\
		\cline{2-2}
		&\makecell[l]{\textbf{Triples}:\\ \hspace{1em}\textbf{Projection}:\\ \hspace{2em}(IBM, considered\_processors, ?y)\\\hspace{1em}\textbf{Negation}:\\ \hspace{2em}$\neg$(IBM PC, used\_processor, ?y) \\\hspace{1em}\textbf{Intersection}:\\ \hspace{2em}(IBM, considered\_processors, ?y) $\land$ \\ \hspace{2em} $\neg$(IBM PC, used\_processor, ?y)} & \\
		\hline
	\end{tabular}
    \caption{Illustrations of structure conformity.}\label{tab2}
\end{table}

\subsubsection{Subgraph-Guided Prompt Design}
To automatically acquire queries with specific logical structures, we design a subgraph-guided prompt that enables LLMs to generate queries based on selected passages. Specifically, LLMs often struggle to internalize and reproduce complex reasoning patterns (such as projection, intersection, and negation) by relying solely on natural language descriptions. To address this, we incorporate symbolic subgraph patterns into the prompt, allowing the LLM to align natural language queries with corresponding logical structures. As shown in Figure~\ref{figure3}, a full prompt consists of three components: \emph{query definition}, \emph{subgraph indicator}, and \emph{demonstration}, where the \emph{subgraph indicator} guides the LLM to learn subgraph patterns associated with different query types, enabling consistent query generation. This design combines symbolic logic for structural control with LLMs' strengths in natural language generation, enabling LLMs to generate queries with complex logical structures. For example, given a \emph{pni} query ``Which actors starred in a film directed by Christopher Nolan but have never won an Oscar?", the corresponding \emph{subgraph indicator} is formulated as:
\[
\{?z \mid (?x, R1, ?y) \wedge (?y, R2, ?z)\} \cap \{?z \mid \neg(?w, R3, ?z)\},
\]
where $?x$ and $?y$ denote the starting constant entities \textit{Christopher} and \textit{Oscar}, respectively. In addition, $?w$ is an intermediate variable representing \textit{film} entities, and $?z$ refers to the target variable entities (\textit{actors}). $R1$, $R2$, and $R3$ denote the respective relations between entities. The prompt defines a projection chain and a negation constraint, which are combined via a set intersection to identify the set of actors. All prompt examples are provided in Appendix J.
\begin{table*}[t]
        \footnotesize
	\centering
	
	\begin{tabular}{lccccccccccccccc}
		\toprule
		Models & 1p & 2p & 3p & 2i & 3i & pi & ip & 2u & up & 2in & 3in & inp & pin & pni &total\\
		\midrule
             HyDE & 65.8 & 56.7 & 52.1 & 58.4 & 57.1 & 44.7 & 48.3 & 49.7 & 45.2 & 36.5 & 31.8 & 38.3 & 31.7 & 34.0 & 45.6\\
             BGE & 66.3 & 56.6 & 53.9 & 60.1 & 58.2 & 45.7& 48.5 & 49.5 & 43.4 & 37.9 & 31.3 & 38.9 & 33.3 & 34.8 & 47.4\\
             PromptReps& 64.6 & 63.4 & 58.0 & 61.4 & 57.4 & 47.7 & 47.8 & 53.5 & 46.7 & 35.7 & 30.2 & 37.6 & 35.7 & 29.6 & 48.6\\
		BM25& 66.1 & 60.2 & 57.4 & 63.5 & 60.3 & 46.2 & 51.6 & 52.7 & \underline{39.5} & 39.3 & 32.2 & 36.6 & 32.4 & 31.7 & 50.5\\
        LameR & 69.6 & 58.8 & 56.9 & \textbf{65.9} & 58.0 & 50.6 & 52.1 & 55.7 & 45.7 & 37.0 & 33.0 & 38.1 & \textbf{35.8} & 33.5 & 52.8\\
		Contriever & 70.2 & 65.3 & \textbf{61.7} & 60.7 & 61.2 & \underline{52.0} & 54.8 & 57.4 & 48.8 & 36.9 & 33.2 & 38.3 & 32.1 & \underline{35.5} & 53.4\\
        AGR& \textbf{74.3} & \underline{61.2} & \underline{60.3} & 62.3 & \textbf{62.7} & 48.4 & \underline{53.0} & \textbf{59.1} & \textbf{52.4} & 35.5 & \underline{34.5} & \textbf{42.3} & \underline{35.5} & 33.8 & \underline{54.3}\\
            InteR& \underline{71.8} & \textbf{64.4} & 58.7 & \underline{63.6} & \underline{62.6} & \textbf{52.3} & \textbf{55.8} & \underline{58.5} & \underline{50.3} & \textbf{39.3} & \textbf{34.7} & \underline{40.3} & 34.6 & \textbf{37.5} & \textbf{55.7}\\
		\bottomrule
	\end{tabular}
    \caption{All retrieval models' nDCG@10 (\%) scores on ComLQ queries across all query types.}\label{tab3}
\end{table*}
\subsubsection{Expert Verification}
To ensure the quality of generated queries, we apply expert verification to all query-passage pairs. Although LLMs can generate fluent outputs, they often exhibit structural hallucinations, i.e., produce queries deviating from the intended structure, or rely on evidence not properly distributed across the supporting passages. To address this problem, we also provide auxiliary query triplets and supporting evidence to assist three experienced annotators in reviewing each generated query-passage pair, concerning \emph{structure conformity} and \emph{evidence distribution} as follows.

\begin{itemize}
\item \textbf{Structure Conformity} As shown in Table~\ref{tab2}, we provide auxiliary query triplets to assist filtering low-quality samples. Three annotators manually review all queries and their corresponding triple forms to validate strict adherence to the target query structure. For example, the positive query ``\emph{Which actor starred in a film directed by Steven Spielberg but not an American?}" conforms to pin structure, while the negative query ``\emph{Which microprocessors were considered for the original IBM PC but were not used in its final design?}" is not a strict pin query. Furthermore, we use majority voting to achieve a consensus for ambiguous queries. This two-step process helps minimize false positives (well-formed queries are incorrectly rejected) and false negatives (ill-formed queries are incorrectly accepted). 

\item \textbf{Evidence Distribution} In addition, we assess the evidence distribution of generated samples. In other words, for queries generated from multiple passages, the annotators verify whether necessary supporting evidence is indeed distributed across those passages. If the evidence is not properly distributed, we discard the corresponding query-passage pair. This annotation stage follows the same protocol as structure conformity, including majority voting and a random sample check.
\end{itemize}

Furthermore, to assess the model's ability to disregard irrelevant information, we augment the corpus with distractor passages whose titles are entirely disjoint from those of the sampled passages, ensuring they are completely unrelated to the generated queries.

\section{Experiments}

\subsection{Retrieval Models}
We first consider several zero-shot retrieval models not involving query-document relevance labels in our experiments, including sparse retriever BM25 \cite{INR-019}, dense retriever BGE \cite{10.1145/3626772.3657878}, and Contriever \cite{2021arXiv211209118I}. Besides, we consider some LLM-based retrieval models, including HyDE \cite{gao-etal-2023-precise}, InteR \cite{feng-etal-2024-synergistic}, LameR \cite{shen-etal-2024-retrieval}, AGR \cite{chen-etal-2024-analyze}, and PromptReps \cite{zhuang-etal-2024-promptreps}.
\subsection{Implementation Details}
To ensure a fair and consistent comparison, we reproduce results on HyDE, LameR, AGR, and InteR, using \texttt{bge-small-en-v1.5} as the embedding model and \texttt{GPT-4o} with a
temperature of 0.5 as the underlying LLM. For PromptReps, we adopt \texttt{LLaMA3-70B-Instruct} to generate hybrid representations. All experiments are conducted on three \texttt{NVIDIA A800 80GB} GPUs.
\subsection{Evaluation Metrics}
In our experiment results, we report the nDCG@10 scores of all retrieval models on ComLQ queries, given nDCG's robustness in capturing the effectiveness of IR models across tasks with binary and graded relevance judgments. Especially, to evaluate the ability of retrievers to handle queries with negation in ComLQ, we propose a novel evaluation metric, \textbf{Log-Scaled Negation Consistency (LSNC@$K$)}, which measures whether the Top-$K$ retrieved passages violate negation conditions specified in queries. Unlike standard relevance-based metrics (e.g., nDCG and mAP) measuring overall relevance, LSNC targets negation consistency evaluation, which is a credible metric for evaluating IR performance on queries with negation. Formally, let $\mathcal{D}_k$ denote the set of top-$K$ retrieved passages,
its LSNC@\emph{K} score is computed as
\begin{equation}
    \text{LSNC@\emph{K}} = -\frac{\log\bigg( \big(\sum\limits_{d\in \mathcal{D}_k}V(d)+1\big)/(K + 1) \bigg)}{\log(K + 1)},
\end{equation}
where $V(d) \in \{0, 1\}$ is an indicator function, and equals 1 if the retrieved passage $d$ violates negation conditions, otherwise 0.
A higher LSNC@$K$ score indicates there are fewer violations in the top-$K$ passages. 

\subsection{Experiment Results}
\subsubsection{Overall Results}\label{sec4.3}
Table~\ref{tab3} presents the nDCG@10 scores of all retrieval models across 14 query types, along with the overall performance in \emph{total} column, where the best scores are highlighted in bold and second-best scores are underlined.
The results show that none of the retrieval models can consistently achieve the best performance across all query types, highlighting the importance of developing specialized retrieval approaches tailored to the unique reasoning demands of various logical structures. In addition, there are some significant findings from the results as follows.

\noindent 1. \textbf{All retrievers' performance degrades as query complexity increases}. The trend of performance degradation is observed in the 
results across \emph{1p}, \emph{2p}, and \emph{3p} queries, demonstrating the inherent difficulty of multi-hop queries. A similar trend is also observed in the results of other query types (\emph{2i} vs. \emph{3i} and \emph{2in} vs. \emph{3in}). Besides, the retrievers perform better on \emph{2u} queries compared to \emph{up} queries, as the latter introduces an additional projection operation that complicates reasoning. The performance gaps reveal the limited capabilities of the retrievers in capturing and reasoning over increasingly intricate logical structures. 

\noindent 2. \textbf{All retrievers perform poorly on queries with negation}. Their performance on queries with negation (\emph{2in}, \emph{3in}, \emph{inp}, \emph{pin}, and \emph{pni}) is consistently lower than that on queries without negation. This reveals that current retrieval models struggle to handle the reasoning scenarios requiring the exclusion of specific entities or relations. The performance gap aligns with prior research \cite{xu-etal-2025-logical}, suggesting that relying on word co-occurrence makes retrievers particularly ill-suited for queries with negation. 

\noindent 3. \textbf{Sparse retrievers remain competitive with dense retrievers}.
BM25 consistently outperforms both BGE and HyDE across all query types. These results violate the common assumption that dense retrievers universally outperform sparse ones,  but conform to the observations from the BEIR benchmark \cite{NEURIPS-DATASETS-AND-BENCHMARKS2021_65b9eea6} where traditional sparse methods maintain robust performance across diverse datasets.

\begin{figure}[t]
    \centering
    \includegraphics[width=\linewidth]{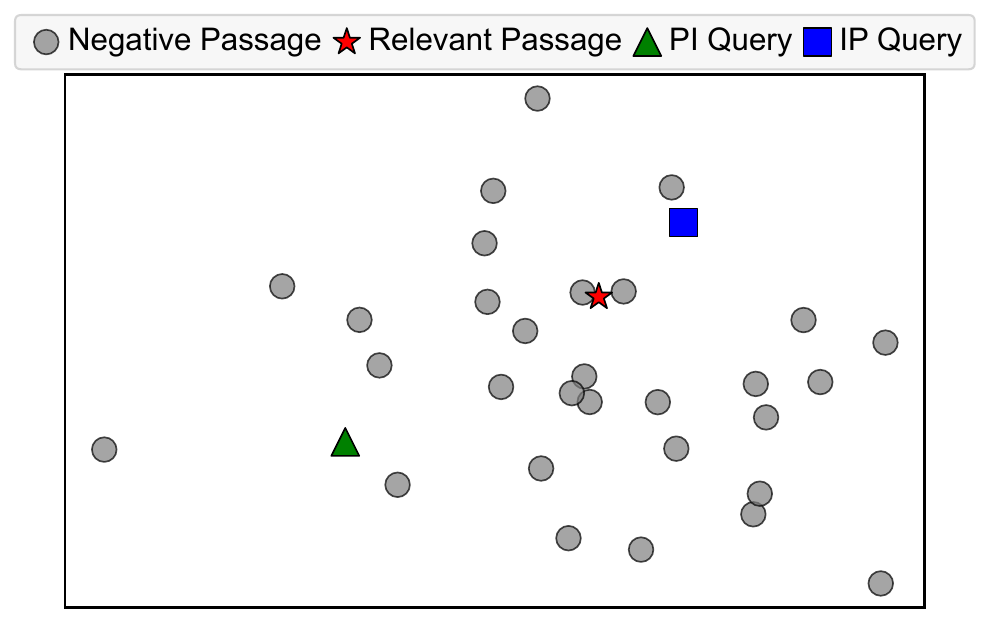}
    \caption{A visualized illustration of query and passage embeddings from ComLQ.}\label{figure4}
\end{figure}

\begin{figure}[t]
    \centering
    \includegraphics[width=\linewidth]{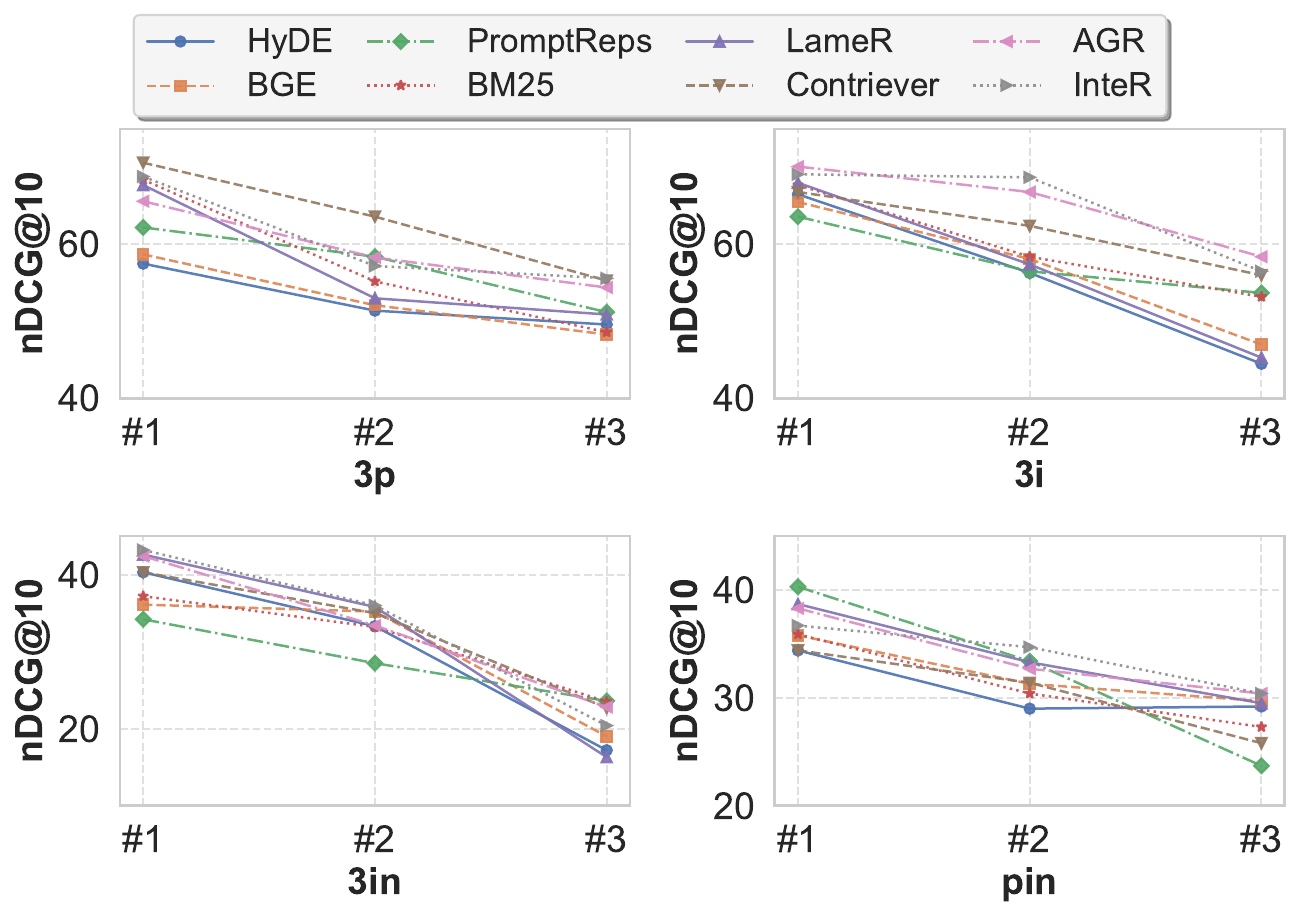}
    \caption{All retrievers' performance with varying numbers of supporting passages on queries of \emph{3p}, \emph{3i}, \emph{3in} and \emph{pin}, respectively.}\label{figure5}
\end{figure}

\begin{table}[htbp]
\centering
\footnotesize

\begin{tabular}{l c c c c c}
\toprule 
Models & 2in & 3in & inp & pin & pni \\
\midrule
HyDE & 27.8 & 26.0 & 23.1 & 24.4 & 24.9 \\
BGE & 30.2 & 31.8 & 26.6 & 27.2 & 25.7 \\
PromptReps  & 31.3 & 33.4 & 27.4 & 26.8 & 29.7 \\
BM25   & 32.2 & 29.0 & 30.5 & 29.3 & 27.4 \\
LameR & 35.6 & 34.7 & 29.9 & 27.2 & 26.4 \\
Contriever & 34.4 & 37.4 & 28.1 & 30.4 & 31.2 \\
AGR & \underline{36.4} & \underline{37.3} & \underline{31.2} & \textbf{32.8} & \textbf{32.2} \\
InteR & \textbf{38.4} & \textbf{38.3} & \textbf{33.2} & \underline{31.7} & \underline{31.7} \\
\bottomrule
\end{tabular}
\caption{LSNC@100 results (\%) on ComLQ across queries with negation.}\label{tab4}
\end{table}

\begin{table*}[htbp]
\centering
\footnotesize

\begin{tabularx}{\textwidth}{|p{1cm}|X|}
\hline
\multicolumn{2}{|l|}{\emph{Pin} query: Which actor starred in a film directed by Steven Spielberg but not an American?}\\
\hline
\textbf{Models} & \textbf{Rewriting queries} \\
\hline
\multirow{3}{*}{HyDE} & Steven Spielberg, one of the most influential \textcolor{red}{American} directors... One notable non-\textcolor{red}{American} actor who starred in a Spielberg-directed film is Daniel Day-Lewis... Although \textit{Lincoln} is a deeply American story, Day-Lewis himself is not an \textcolor{red}{American} citizen, making him a prominent example of a non-\textcolor{red}{American} actor starring in a Spielberg film. \\
\hline
\multirow{2}{*}{InteR} & Which actor starred in a film directed by Steven Spielberg but not an \textcolor{red}{American}?...One notable non-\textcolor{red}{American} actor who starred in a Spielberg-directed film is Ralph Fiennes...while not being an \textcolor{red}{American}, making him a correct answer to the question. \\
\hline
\multirow{3}{*}{LameR} & Which actor starred in a film directed by Steven Spielberg but not an \textcolor{red}{American}? Steven Spielberg has worked with many actors from around the world, including several who are not \textcolor{red}{American}...in an \textcolor{red}{American}-directed film. His British nationality makes him a correct answer to the question of which actor starred in a Spielberg film but is not \textcolor{red}{American}. \\
\hline
\multirow{2}{*}{AGR} & Ben Kingsley, a British actor, starred in Steven Spielberg's \textit{Schindler's List} (1993) as Itzhak Stern. Kingsley is not American, making him a correct example of a non-\textcolor{red}{American} actor who starred in a Spielberg-directed film. \\
\hline
\end{tabularx}
\caption{Query examples rewritten by LLM-based query rewriting models HyDE, InteR, LameR and AGR, respectively.}\label{tab5}

\end{table*}
 
\subsubsection{Impact of Logical Operation Order}
From the experiment results, we also observe that the retrievers exhibit inferior retrieval performance on projection-then-intersection queries than intersection-then-projection queries (\emph{pi} vs. \emph{ip}, \emph{pin} and \emph{pni} vs. \emph{inp}). To illustrate it, we compare a \emph{pi} query example (``\emph{Find people who acted in a movie directed by Christopher Nolan and who also won an Oscar}") with an \emph{ip} query example (``\emph{Which universities have produced individuals who are both Nobel Prize winners and Fields Medalists?}") from ComLQ. The embeddings of the two queries and their corresponding passages are visualized using t-SNE in Figure~\ref{figure4}. Although two queries have the same relevant passage, the results show that the embedding of \emph{ip} query is closer to that of the passage than the \emph{pi} query. We argue that projection-then-intersection queries generally involve more complex semantic compositions, which tend to confuse retrievers and lead queries to less alignment with relevant passages. 
 
\subsubsection{Impact of Evidence Distribution}
To investigate the impact of the number of supporting passages on retrievers' performance, in Figure~\ref{figure5}, we depict all retrievers' performance lines when varying evidence distribution levels (we only display the results on the representative query types \emph{3p}, \emph{3i}, \emph{3in} and \emph{pin} due to space limitation). Specifically, \#1, \#2 and \#3 denote the cases where the correct answer must be inferred from one, two or three supporting passages, respectively. The results show that queries requiring evidence from more passages are more challenging for retrievers.  

\subsubsection{Performance on Negation Queries}
To evaluate whether retrievers can correctly handle queries with negation in ComLQ, we evaluate retrieval models with our proposed metric LSNC@$K$. As shown in Table~\ref{tab4}, all retrievers exhibit very low LSNC@100 scores on the five query types with negation, i.e., \emph{2in}, \emph{3in}, \emph{inp}, \emph{pin}, and \emph{pni},
and are weaker on queries involving multiple operations (\emph{inp}, \emph{pin} and \emph{pni}). This suggests that increased logical complexity further impairs the models’ ability to handle negation conditions, highlighting the need for retrievers capable of modeling exclusion. The alignment between low LSNC and nDCG scores on queries with negation further reveals that retrievers tend to prioritize passages containing negation conditions, thus resulting in poor overall relevance.
\subsection{Effect of Subgraph Indicator}
\begin{figure}[t]
    \centering
    \includegraphics[width=0.85\linewidth]{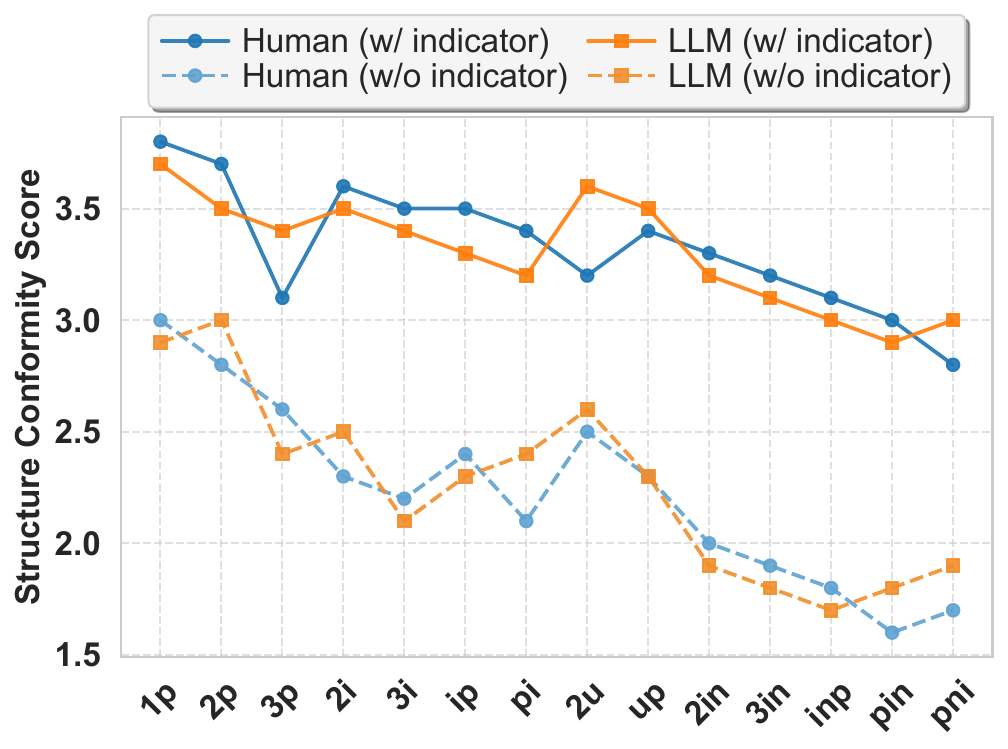}
    \caption{Structure conformity scores of human and LLM assessment across 14 query types with and without the subgraph indicator, respectively.}\label{figure6}
\end{figure}

To evaluate the effect of the subgraph indicator, we conduct an ablation study by removing the corresponding component in the prompt of data generation. Then, the structure conformity of the generated queries is assessed by the human and  LLM (GPT-4o), respectively. The assessment score scale is from 1 to 4, where higher scores indicate better generation qualities. 
Figure~\ref{figure6} reports the average structure conformity scores of human and LLM assessments across 14 query types with and without the subgraph indicator, respectively. Two assessments show that removing the subgraph indicator leads to a drop in structure conformity scores across all query types. Furthermore, the more complex query types (in the right part of Figure~\ref{figure6}) get lower scores, verifying the challenges LLMs encounter when generating more complex queries. These ablation study results highlight the significance of the subgraph indicator in guiding LLMs to generate structurally consistent queries.

\subsection{Case Study}
It has been proven that LLM-based query rewriting models such as HyDE, InteR, LameR, and AGR achieve strong performance on some benchmarks, including BEIR \cite{NEURIPS-DATASETS-AND-BENCHMARKS2021_65b9eea6}, TREC DL'19 \cite{craswell2020overviewtrec2019deep} and DL'20 \cite{DBLP:conf/trec/CraswellMMYC20}. 
However, our empirical studies show that these models exhibit catastrophic failures on  queries with negation.
To illustrate this finding, in Table~\ref{tab5} we display queries rewritten respectively by the four models for the \emph{pin} query. 
It shows that all rewriting queries explicitly retain the negation condition `American'. Such incorrectly rewriting queries cause these word co-occurrence-based retrievers to favor the irrelevant documents containing the term `American', demonstrating their misunderstanding of the original query intent \cite{xu-etal-2025-logical}.


\section{Conclusion}
In this paper, we introduce \textbf{ComLQ}, a novel dataset to evaluate IR systems on complex logical queries, which are overlooked by existing IR benchmarks. We propose an LLM-based data synthesis method to construct ComLQ, which consists of passage selection, data generation by the subgraph-guided prompt with a subgraph indicator, expert verification for structure conformity and evidence distribution. Our experiment results under zero-shot settings demonstrate that existing retrieval models exhibit significant limitations on complex logical queries. Our findings also emphasize the need for retrieval models capable of reasoning over complex structures. To further measure how well retrievers handle negation conditions, we further propose a new evaluation metric, LSNC, of which the scores reveal consistently low negation consistency across all retrieval models on queries with negation.
\section{Acknowledgments}
The authors disclosed receipt of the following financial support for the research, authorship, and publication of this article: This research was supported by the Chinese NSF Major Research Plan (No.92270121), General Program (No.62572129) and the AI Laboratory of United Automotive Electronic Systems (UAES) Co. (Grant no. 2025-3944).
\bibliography{aaai2026}

@inproceedings{10.1145/3637528.3671458,
author = {Dai, Sunhao and Xu, Chen and Xu, Shicheng and Pang, Liang and Dong, Zhenhua and Xu, Jun},
title = {Bias and Unfairness in Information Retrieval Systems: New Challenges in the LLM Era},
year = {2024},
isbn = {9798400704901},
publisher = {Association for Computing Machinery},
address = {New York, NY, USA},
url = {https://doi.org/10.1145/3637528.3671458},
doi = {10.1145/3637528.3671458},
booktitle = {Proceedings of the 30th ACM SIGKDD Conference on Knowledge Discovery and Data Mining},
pages = {6437–6447},
numpages = {11},
location = {Barcelona, Spain},
series = {KDD '24}
}

@inproceedings{karpukhin-etal-2020-dense,
    title = "Dense Passage Retrieval for Open-Domain Question Answering",
    author = "Karpukhin, Vladimir  and
      Oguz, Barlas  and
      Min, Sewon  and
      Lewis, Patrick  and
      Wu, Ledell  and
      Edunov, Sergey  and
      Chen, Danqi  and
      Yih, Wen-tau",
    editor = "Webber, Bonnie  and
      Cohn, Trevor  and
      He, Yulan  and
      Liu, Yang",
    booktitle = "Proceedings of the 2020 Conference on Empirical Methods in Natural Language Processing (EMNLP)",
    month = nov,
    year = "2020",
    address = "Online",
    publisher = "Association for Computational Linguistics",
    url = "https://aclanthology.org/2020.emnlp-main.550/",
    doi = "10.18653/v1/2020.emnlp-main.550",
    pages = "6769--6781"
}

@article{10.1145/3637870,
author = {Zhao, Wayne Xin and Liu, Jing and Ren, Ruiyang and Wen, Ji-Rong},
title = {Dense Text Retrieval Based on Pretrained Language Models: A Survey},
year = {2024},
issue_date = {July 2024},
publisher = {Association for Computing Machinery},
address = {New York, NY, USA},
volume = {42},
number = {4},
issn = {1046-8188},
url = {https://doi.org/10.1145/3637870},
doi = {10.1145/3637870},
journal = {ACM Trans. Inf. Syst.},
month = feb,
articleno = {89},
numpages = {60}
}

@inproceedings{DBLP:conf/nips/NguyenRSGTMD16,
  author       = {Tri Nguyen and
                  Mir Rosenberg and
                  Xia Song and
                  Jianfeng Gao and
                  Saurabh Tiwary and
                  Rangan Majumder and
                  Li Deng},
  editor       = {Tarek Richard Besold and
                  Antoine Bordes and
                  Artur S. d'Avila Garcez and
                  Greg Wayne},
  title        = {{MS} {MARCO:} {A} Human Generated MAchine Reading COmprehension Dataset},
  booktitle    = {Proceedings of the Workshop on Cognitive Computation: Integrating
                  neural and symbolic approaches 2016 co-located with the 30th Annual
                  Conference on Neural Information Processing Systems {(NIPS} 2016),
                  Barcelona, Spain, December 9, 2016},
  series       = {{CEUR} Workshop Proceedings},
  volume       = {1773},
  publisher    = {CEUR-WS.org},
  year         = {2016},
  url          = {https://ceur-ws.org/Vol-1773/CoCoNIPS\_2016\_paper9.pdf},
  bibsource    = {dblp computer science bibliography, https://dblp.org}
}

@inproceedings{DBLP:conf/trec/CraswellMMYC20,
  author       = {Nick Craswell and
                  Bhaskar Mitra and
                  Emine Yilmaz and
                  Daniel Campos},
  editor       = {Ellen M. Voorhees and
                  Angela Ellis},
  title        = {Overview of the {TREC} 2020 Deep Learning Track},
  booktitle    = {Proceedings of the Twenty-Ninth Text REtrieval Conference, {TREC}
                  2020, Virtual Event [Gaithersburg, Maryland, USA], November 16-20,
                  2020},
  series       = {{NIST} Special Publication},
  volume       = {1266},
  publisher    = {National Institute of Standards and Technology {(NIST)}},
  year         = {2020},
  url          = {https://trec.nist.gov/pubs/trec29/papers/OVERVIEW.DL.pdf},
  bibsource    = {dblp computer science bibliography, https://dblp.org}
}

@inproceedings{NEURIPS2020_e43739bb,
 author = {Ren, Hongyu and Leskovec, Jure},
 booktitle = {Advances in Neural Information Processing Systems},
 editor = {H. Larochelle and M. Ranzato and R. Hadsell and M.F. Balcan and H. Lin},
 pages = {19716--19726},
 publisher = {Curran Associates, Inc.},
 title = {Beta Embeddings for Multi-Hop Logical Reasoning in Knowledge Graphs},
 url = {https://proceedings.neurips.cc/paper_files/paper/2020/file/e43739bba7cdb577e9e3e4e42447f5a5-Paper.pdf},
 volume = {33},
 year = {2020}
}

@inproceedings{DBLP:conf/iclr/RenHL20,
  author       = {Hongyu Ren and
                  Weihua Hu and
                  Jure Leskovec},
  title        = {Query2box: Reasoning over Knowledge Graphs in Vector Space Using Box
                  Embeddings},
  booktitle    = {8th International Conference on Learning Representations, {ICLR} 2020,
                  Addis Ababa, Ethiopia, April 26-30, 2020},
  publisher    = {OpenReview.net},
  year         = {2020},
  url          = {https://openreview.net/forum?id=BJgr4kSFDS},
  bibsource    = {dblp computer science bibliography, https://dblp.org}
}

@inproceedings{fang-etal-2024-complex,
    title = "Complex Reasoning over Logical Queries on Commonsense Knowledge Graphs",
    author = "Fang, Tianqing  and
      Chen, Zeming  and
      Song, Yangqiu  and
      Bosselut, Antoine",
    editor = "Ku, Lun-Wei  and
      Martins, Andre  and
      Srikumar, Vivek",
    booktitle = "Proceedings of the 62nd Annual Meeting of the Association for Computational Linguistics (Volume 1: Long Papers)",
    month = aug,
    year = "2024",
    address = "Bangkok, Thailand",
    publisher = "Association for Computational Linguistics",
    url = "https://aclanthology.org/2024.acl-long.613/",
    doi = "10.18653/v1/2024.acl-long.613",
    pages = "11365--11384"
}

@inproceedings{ji-etal-2024-retrieval,
    title = "Retrieval and Reasoning on {KG}s: Integrate Knowledge Graphs into Large Language Models for Complex Question Answering",
    author = "Ji, Yixin  and
      Wu, Kaixin  and
      Li, Juntao  and
      Chen, Wei  and
      Zhong, Mingjie  and
      Jia, Xu  and
      Zhang, Min",
    editor = "Al-Onaizan, Yaser  and
      Bansal, Mohit  and
      Chen, Yun-Nung",
    booktitle = "Findings of the Association for Computational Linguistics: EMNLP 2024",
    month = nov,
    year = "2024",
    address = "Miami, Florida, USA",
    publisher = "Association for Computational Linguistics",
    url = "https://aclanthology.org/2024.findings-emnlp.446/",
    doi = "10.18653/v1/2024.findings-emnlp.446",
    pages = "7598--7610"
}

@article{INR-019,
url = {http://dx.doi.org/10.1561/1500000019},
year = {2009},
volume = {3},
journal = {Foundations and Trends® in Information Retrieval},
title = {The Probabilistic Relevance Framework: BM25 and Beyond},
doi = {10.1561/1500000019},
issn = {1554-0669},
number = {4},
pages = {333-389},
author = {Stephen Robertson and Hugo Zaragoza}
}

@inproceedings{10.1145/3626772.3657878,
author = {Xiao, Shitao and Liu, Zheng and Zhang, Peitian and Muennighoff, Niklas and Lian, Defu and Nie, Jian-Yun},
title = {C-Pack: Packed Resources For General Chinese Embeddings},
year = {2024},
isbn = {9798400704314},
publisher = {Association for Computing Machinery},
address = {New York, NY, USA},
url = {https://doi.org/10.1145/3626772.3657878},
doi = {10.1145/3626772.3657878},
booktitle = {Proceedings of the 47th International ACM SIGIR Conference on Research and Development in Information Retrieval},
pages = {641–649},
numpages = {9},
location = {Washington DC, USA},
series = {SIGIR '24}
}

@ARTICLE{2021arXiv211209118I,
       author = {{Izacard}, Gautier and {Caron}, Mathilde and {Hosseini}, Lucas and {Riedel}, Sebastian and {Bojanowski}, Piotr and {Joulin}, Armand and {Grave}, Edouard},
        title = "{Unsupervised Dense Information Retrieval with Contrastive Learning}",
      journal = {arXiv e-prints},
         year = 2021,
        month = dec,
          eid = {arXiv:2112.09118},
        pages = {arXiv:2112.09118},
          doi = {10.48550/arXiv.2112.09118},
archivePrefix = {arXiv},
       eprint = {2112.09118},
 primaryClass = {cs.IR},
       adsurl = {https://ui.adsabs.harvard.edu/abs/2021arXiv211209118I}
}

@inproceedings{gao-etal-2023-precise,
    title = "Precise Zero-Shot Dense Retrieval without Relevance Labels",
    author = "Gao, Luyu  and
      Ma, Xueguang  and
      Lin, Jimmy  and
      Callan, Jamie",
    editor = "Rogers, Anna  and
      Boyd-Graber, Jordan  and
      Okazaki, Naoaki",
    booktitle = "Proceedings of the 61st Annual Meeting of the Association for Computational Linguistics (Volume 1: Long Papers)",
    month = jul,
    year = "2023",
    address = "Toronto, Canada",
    publisher = "Association for Computational Linguistics",
    url = "https://aclanthology.org/2023.acl-long.99/",
    doi = "10.18653/v1/2023.acl-long.99",
    pages = "1762--1777"
}

@inproceedings{feng-etal-2024-synergistic,
    title = "Synergistic Interplay between Search and Large Language Models for Information Retrieval",
    author = "Feng, Jiazhan  and
      Tao, Chongyang  and
      Geng, Xiubo  and
      Shen, Tao  and
      Xu, Can  and
      Long, Guodong  and
      Zhao, Dongyan  and
      Jiang, Daxin",
    editor = "Ku, Lun-Wei  and
      Martins, Andre  and
      Srikumar, Vivek",
    booktitle = "Proceedings of the 62nd Annual Meeting of the Association for Computational Linguistics (Volume 1: Long Papers)",
    month = aug,
    year = "2024",
    address = "Bangkok, Thailand",
    publisher = "Association for Computational Linguistics",
    url = "https://aclanthology.org/2024.acl-long.517/",
    doi = "10.18653/v1/2024.acl-long.517",
    pages = "9571--9583"
}

@inproceedings{zhuang-etal-2024-promptreps,
    title = "{P}rompt{R}eps: Prompting Large Language Models to Generate Dense and Sparse Representations for Zero-Shot Document Retrieval",
    author = "Zhuang, Shengyao  and
      Ma, Xueguang  and
      Koopman, Bevan  and
      Lin, Jimmy  and
      Zuccon, Guido",
    editor = "Al-Onaizan, Yaser  and
      Bansal, Mohit  and
      Chen, Yun-Nung",
    booktitle = "Proceedings of the 2024 Conference on Empirical Methods in Natural Language Processing",
    month = nov,
    year = "2024",
    address = "Miami, Florida, USA",
    publisher = "Association for Computational Linguistics",
    url = "https://aclanthology.org/2024.emnlp-main.250/",
    doi = "10.18653/v1/2024.emnlp-main.250",
    pages = "4375--4391"
}

@inproceedings{shen-etal-2024-retrieval,
    title = "Retrieval-Augmented Retrieval: Large Language Models are Strong Zero-Shot Retriever",
    author = "Shen, Tao  and
      Long, Guodong  and
      Geng, Xiubo  and
      Tao, Chongyang  and
      Lei, Yibin  and
      Zhou, Tianyi  and
      Blumenstein, Michael  and
      Jiang, Daxin",
    editor = "Ku, Lun-Wei  and
      Martins, Andre  and
      Srikumar, Vivek",
    booktitle = "Findings of the Association for Computational Linguistics: ACL 2024",
    month = aug,
    year = "2024",
    address = "Bangkok, Thailand",
    publisher = "Association for Computational Linguistics",
    url = "https://aclanthology.org/2024.findings-acl.943/",
    doi = "10.18653/v1/2024.findings-acl.943",
    pages = "15933--15946"
}

@inproceedings{chen-etal-2024-analyze,
    title = "Analyze, Generate and Refine: Query Expansion with {LLM}s for Zero-Shot Open-Domain {QA}",
    author = "Chen, Xinran  and
      Chen, Xuanang  and
      He, Ben  and
      Wen, Tengfei  and
      Sun, Le",
    editor = "Ku, Lun-Wei  and
      Martins, Andre  and
      Srikumar, Vivek",
    booktitle = "Findings of the Association for Computational Linguistics: ACL 2024",
    month = aug,
    year = "2024",
    address = "Bangkok, Thailand",
    publisher = "Association for Computational Linguistics",
    url = "https://aclanthology.org/2024.findings-acl.708/",
    doi = "10.18653/v1/2024.findings-acl.708",
    pages = "11908--11922"
}

@inproceedings{NEURIPS-DATASETS-AND-BENCHMARKS2021_65b9eea6,
 author = {Thakur, Nandan and Reimers, Nils and R\"{u}ckl\'{e}, Andreas and Srivastava, Abhishek and Gurevych, Iryna},
 booktitle = {Proceedings of the Neural Information Processing Systems Track on Datasets and Benchmarks},
 editor = {J. Vanschoren and S. Yeung},
 pages = {},
 title = {BEIR: A Heterogeneous Benchmark for Zero-shot Evaluation of Information Retrieval Models},
 url = {https://datasets-benchmarks-proceedings.neurips.cc/paper_files/paper/2021/file/65b9eea6e1cc6bb9f0cd2a47751a186f-Paper-round2.pdf},
 volume = {1},
 year = {2021}
}

@inproceedings{NEURIPS2023_6174c67b,
 author = {Bai, Jiaxin and Liu, Xin and Wang, Weiqi and Luo, Chen and Song, Yangqiu},
 booktitle = {Advances in Neural Information Processing Systems},
 editor = {A. Oh and T. Naumann and A. Globerson and K. Saenko and M. Hardt and S. Levine},
 pages = {30534--30553},
 publisher = {Curran Associates, Inc.},
 title = {Complex Query Answering on Eventuality Knowledge Graph with Implicit Logical Constraints},
 url = {https://proceedings.neurips.cc/paper_files/paper/2023/file/6174c67b136621f3f2e4a6b1d3286f6b-Paper-Conference.pdf},
 volume = {36},
 year = {2023}
}

@article{LIN2023253,
title = {Fusing topology contexts and logical rules in language models for knowledge graph completion},
journal = {Information Fusion},
volume = {90},
pages = {253-264},
year = {2023},
issn = {1566-2535},
doi = {https://doi.org/10.1016/j.inffus.2022.09.020},
url = {https://www.sciencedirect.com/science/article/pii/S1566253522001592},
author = {Qika Lin and Rui Mao and Jun Liu and Fangzhi Xu and Erik Cambria}
}

@inproceedings{NEURIPS2018_ef50c335,
 author = {Hamilton, Will and Bajaj, Payal and Zitnik, Marinka and Jurafsky, Dan and Leskovec, Jure},
 booktitle = {Advances in Neural Information Processing Systems},
 editor = {S. Bengio and H. Wallach and H. Larochelle and K. Grauman and N. Cesa-Bianchi and R. Garnett},
 pages = {},
 publisher = {Curran Associates, Inc.},
 title = {Embedding Logical Queries on Knowledge Graphs},
 url = {https://proceedings.neurips.cc/paper_files/paper/2018/file/ef50c335cca9f340bde656363ebd02fd-Paper.pdf},
 volume = {31},
 year = {2018}
}

@inproceedings{fangetal2024complex,
    title = "Complex Reasoning over Logical Queries on Commonsense Knowledge Graphs",
    author = "Fang, Tianqing  and
      Chen, Zeming  and
      Song, Yangqiu  and
      Bosselut, Antoine",
    editor = "Ku, Lun-Wei  and
      Martins, Andre  and
      Srikumar, Vivek",
    booktitle = "Proceedings of the 62nd Annual Meeting of the Association for Computational Linguistics (Volume 1: Long Papers)",
    month = aug,
    year = "2024",
    address = "Bangkok, Thailand",
    publisher = "Association for Computational Linguistics",
    url = "https://aclanthology.org/2024.acl-long.613/",
    doi = "10.18653/v1/2024.acl-long.613",
    pages = "11365--11384"
}

@inproceedings{10.1145/3637528.3671869,
author = {Zhang, Chongzhi and Peng, Zhiping and Zheng, Junhao and Ma, Qianli},
title = {Conditional Logical Message Passing Transformer for Complex Query Answering},
year = {2024},
isbn = {9798400704901},
publisher = {Association for Computing Machinery},
address = {New York, NY, USA},
url = {https://doi.org/10.1145/3637528.3671869},
doi = {10.1145/3637528.3671869},
booktitle = {Proceedings of the 30th ACM SIGKDD Conference on Knowledge Discovery and Data Mining},
pages = {4119–4130},
numpages = {12},
location = {Barcelona, Spain},
series = {KDD '24}
}

@inproceedings{leeetal2019latent,
    title = "Latent Retrieval for Weakly Supervised Open Domain Question Answering",
    author = "Lee, Kenton  and
      Chang, Ming-Wei  and
      Toutanova, Kristina",
    editor = "Korhonen, Anna  and
      Traum, David  and
      M{\`a}rquez, Llu{\'i}s",
    booktitle = "Proceedings of the 57th Annual Meeting of the Association for Computational Linguistics",
    month = jul,
    year = "2019",
    address = "Florence, Italy",
    publisher = "Association for Computational Linguistics",
    url = "https://aclanthology.org/P19-1612/",
    doi = "10.18653/v1/P19-1612",
    pages = "6086--6096"
}

@misc{bajaj2018msmarcohumangenerated,
      title={MS MARCO: A Human Generated MAchine Reading COmprehension Dataset}, 
      author={Payal Bajaj and Daniel Campos and Nick Craswell and Li Deng and Jianfeng Gao and Xiaodong Liu and Rangan Majumder and Andrew McNamara and Bhaskar Mitra and Tri Nguyen and Mir Rosenberg and Xia Song and Alina Stoica and Saurabh Tiwary and Tong Wang},
      year={2018},
      eprint={1611.09268},
      archivePrefix={arXiv},
      primaryClass={cs.CL},
      url={https://arxiv.org/abs/1611.09268}, 
}

@inproceedings{10.1145/345508.345577,
author = {Voorhees, Ellen M. and Tice, Dawn M.},
title = {Building a question answering test collection},
year = {2000},
isbn = {1581132263},
publisher = {Association for Computing Machinery},
address = {New York, NY, USA},
url = {https://doi.org/10.1145/345508.345577},
doi = {10.1145/345508.345577},
booktitle = {Proceedings of the 23rd Annual International ACM SIGIR Conference on Research and Development in Information Retrieval},
pages = {200–207},
numpages = {8},
location = {Athens, Greece},
series = {SIGIR '00}
}

@inproceedings{NEURIPSDATASETSANDBENCHMARKS202165b9eea6,
 author = {Thakur, Nandan and Reimers, Nils and R\"{u}ckl\'{e}, Andreas and Srivastava, Abhishek and Gurevych, Iryna},
 booktitle = {Proceedings of the Neural Information Processing Systems Track on Datasets and Benchmarks},
 editor = {J. Vanschoren and S. Yeung},
 pages = {},
 title = {BEIR: A Heterogeneous Benchmark for Zero-shot Evaluation of Information Retrieval Models},
 url = {https://datasets-benchmarks-proceedings.neurips.cc/paper_files/paper/2021/file/65b9eea6e1cc6bb9f0cd2a47751a186f-Paper-round2.pdf},
 volume = {1},
 year = {2021}
}

@inproceedings{muennighoff-etal-2023-mteb,
    title = "{MTEB}: Massive Text Embedding Benchmark",
    author = "Muennighoff, Niklas  and
      Tazi, Nouamane  and
      Magne, Loic  and
      Reimers, Nils",
    editor = "Vlachos, Andreas  and
      Augenstein, Isabelle",
    booktitle = "Proceedings of the 17th Conference of the European Chapter of the Association for Computational Linguistics",
    month = may,
    year = "2023",
    address = "Dubrovnik, Croatia",
    publisher = "Association for Computational Linguistics",
    url = "https://aclanthology.org/2023.eacl-main.148/",
    doi = "10.18653/v1/2023.eacl-main.148",
    pages = "2014--2037"
}

@inproceedings{schimanski-etal-2024-climretrieve,
    title = "{C}lim{R}etrieve: A Benchmarking Dataset for Information Retrieval from Corporate Climate Disclosures",
    author = "Schimanski, Tobias  and
      Ni, Jingwei  and
      Mart{\'i}n, Roberto Spacey  and
      Ranger, Nicola  and
      Leippold, Markus",
    editor = "Al-Onaizan, Yaser  and
      Bansal, Mohit  and
      Chen, Yun-Nung",
    booktitle = "Proceedings of the 2024 Conference on Empirical Methods in Natural Language Processing",
    month = nov,
    year = "2024",
    address = "Miami, Florida, USA",
    publisher = "Association for Computational Linguistics",
    url = "https://aclanthology.org/2024.emnlp-main.969/",
    doi = "10.18653/v1/2024.emnlp-main.969",
    pages = "17509--17524"
}

@inproceedings{su-etal-2024-stard,
    title = "{STARD}: A {C}hinese Statute Retrieval Dataset Derived from Real-life Queries by Non-professionals",
    author = "Su, Weihang  and
      Hu, Yiran  and
      Xie, Anzhe  and
      Ai, Qingyao  and
      Bing, Quezi  and
      Zheng, Ning  and
      Liu, Yun  and
      Shen, Weixing  and
      Liu, Yiqun",
    editor = "Al-Onaizan, Yaser  and
      Bansal, Mohit  and
      Chen, Yun-Nung",
    booktitle = "Findings of the Association for Computational Linguistics: EMNLP 2024",
    month = nov,
    year = "2024",
    address = "Miami, Florida, USA",
    publisher = "Association for Computational Linguistics",
    url = "https://aclanthology.org/2024.findings-emnlp.625/",
    doi = "10.18653/v1/2024.findings-emnlp.625",
    pages = "10658--10671"
}

@inproceedings{ajith-etal-2024-litsearch,
    title = "{L}it{S}earch: A Retrieval Benchmark for Scientific Literature Search",
    author = "Ajith, Anirudh  and
      Xia, Mengzhou  and
      Chevalier, Alexis  and
      Goyal, Tanya  and
      Chen, Danqi  and
      Gao, Tianyu",
    editor = "Al-Onaizan, Yaser  and
      Bansal, Mohit  and
      Chen, Yun-Nung",
    booktitle = "Proceedings of the 2024 Conference on Empirical Methods in Natural Language Processing",
    month = nov,
    year = "2024",
    address = "Miami, Florida, USA",
    publisher = "Association for Computational Linguistics",
    url = "https://aclanthology.org/2024.emnlp-main.840/",
    doi = "10.18653/v1/2024.emnlp-main.840",
    pages = "15068--15083"
}

@inproceedings{NEURIPS2023_78f9c04b,
 author = {Wang, Jianyou (Andre) and Wang, Kaicheng and Wang, Xiaoyue and Naidu, Prudhviraj and Bergen, Leon and Paturi, Ramamohan},
 booktitle = {Advances in Neural Information Processing Systems},
 editor = {A. Oh and T. Naumann and A. Globerson and K. Saenko and M. Hardt and S. Levine},
 pages = {38404--38419},
 publisher = {Curran Associates, Inc.},
 title = {Scientific Document Retrieval using Multi-level Aspect-based Queries},
 url = {https://proceedings.neurips.cc/paper_files/paper/2023/file/78f9c04bdcb06f1ada3902912d8b64ba-Paper-Datasets_and_Benchmarks.pdf},
 volume = {36},
 year = {2023}
}

@inproceedings{xu-etal-2025-logical,
    title = "Logical Consistency is Vital: Neural-Symbolic Information Retrieval for Negative-Constraint Queries",
    author = "Xu, Ganlin  and
      Zhang, Zhoujia  and
      Mei, Wangyi  and
      Liang, Jiaqing  and
      Lu, Weijia  and
      Zhang, Xiaodong  and
      Yang, Zhifei  and
      Ma, Xiaofeng  and
      Xiao, Yanghua  and
      Yang, Deqing",
    editor = "Che, Wanxiang  and
      Nabende, Joyce  and
      Shutova, Ekaterina  and
      Pilehvar, Mohammad Taher",
    booktitle = "Findings of the Association for Computational Linguistics: ACL 2025",
    month = jul,
    year = "2025",
    address = "Vienna, Austria",
    publisher = "Association for Computational Linguistics",
    url = "https://aclanthology.org/2025.findings-acl.92/",
    pages = "1828--1847",
    ISBN = "979-8-89176-256-5"
}

@misc{craswell2020overviewtrec2019deep,
      title={Overview of the TREC 2019 deep learning track}, 
      author={Nick Craswell and Bhaskar Mitra and Emine Yilmaz and Daniel Campos and Ellen M. Voorhees},
      year={2020},
      eprint={2003.07820},
      archivePrefix={arXiv},
      primaryClass={cs.IR},
      url={https://arxiv.org/abs/2003.07820}, 
}

@inproceedings{dietz2017trec,
  title={TREC Complex Answer Retrieval Overview.},
  author={Dietz, Laura and Verma, Manisha and Radlinski, Filip and Craswell, Nick},
  booktitle={TREC},
  year={2017}
}

@inproceedings{yang-etal-2018-hotpotqa,
    title = "{H}otpot{QA}: A Dataset for Diverse, Explainable Multi-hop Question Answering",
    author = "Yang, Zhilin  and
      Qi, Peng  and
      Zhang, Saizheng  and
      Bengio, Yoshua  and
      Cohen, William  and
      Salakhutdinov, Ruslan  and
      Manning, Christopher D.",
    editor = "Riloff, Ellen  and
      Chiang, David  and
      Hockenmaier, Julia  and
      Tsujii, Jun{'}ichi",
    booktitle = "Proceedings of the 2018 Conference on Empirical Methods in Natural Language Processing",
    month = oct # "-" # nov,
    year = "2018",
    address = "Brussels, Belgium",
    publisher = "Association for Computational Linguistics",
    url = "https://aclanthology.org/D18-1259/",
    doi = "10.18653/v1/D18-1259",
    pages = "2369--2380"
}
fi
\makeatletter\def\@listi{\leftmargin\leftmargini \topsep .5em \parsep .5em \itemsep .5em}
\def\@listii{\leftmargin\leftmarginii \labelwidth\leftmarginii \advance\labelwidth-\labelsep \topsep .4em \parsep .4em \itemsep .4em}
\def\@listiii{\leftmargin\leftmarginiii \labelwidth\leftmarginiii \advance\labelwidth-\labelsep \topsep .4em \parsep .4em \itemsep .4em}\makeatother

\setcounter{secnumdepth}{0}

\newcounter{checksubsection}
\newcounter{checkitem}[checksubsection]

\newcommand{\checksubsection}[1]{%
  \refstepcounter{checksubsection}%
  \paragraph{\arabic{checksubsection}. #1}%
  \setcounter{checkitem}{0}%
}

\newcommand{\checkitem}{%
  \refstepcounter{checkitem}%
  \item[\arabic{checksubsection}.\arabic{checkitem}.]%
}
\newcommand{\question}[2]{\normalcolor\checkitem #1 #2 \color{blue}}
\newcommand{\ifyespoints}[1]{\makebox[0pt][l]{\hspace{-15pt}\normalcolor #1}}

\end{document}